\documentclass[aps,prd,noshowpacs, twocolumn,
floatfix,citeautoscript,preprintnumbers]{revtex4-1}
\usepackage[english]{babel}
\usepackage{amssymb}
\usepackage{amsfonts}
\usepackage{amsmath}
\usepackage{eucal}
\usepackage{graphicx}
\usepackage{epsfig}
\usepackage{slashed}

\usepackage[bookmarks=true,colorlinks,linkcolor=blue,urlcolor=blue,citecolor=red]{hyperref}

\newcommand{\be}{\begin{equation}} \newcommand{\ee}{\end{equation}}
\newcommand{\bea}{\begin{eqnarray}} \newcommand{\eea}{\end{eqnarray}}
\newcommand{\beann}{\begin{eqnarray*}}  \newcommand{\eeann}{\end{eqnarray*}}
\newcommand{\bfig}{\begin{figure}} \newcommand{\efig}{\end{figure}}
\newcommand{\ba}{\begin{array}} \newcommand{\ea}{\end{array}}
\newcommand{\bcen}{\begin{center}} \newcommand{\ecen}{\end{center}}
\newcommand{\btab}{\begin{tabular}} \newcommand{\etab}{\end{tabular}}
\newcommand{\nn}{\nonumber}

\newtheorem{Proposition}{Proposition}[section]

\newtheorem{Theorem}{Theorem}[section]
\newtheorem{Lemma}{Lemma}[section]
\newtheorem{Corrolary}{Corrolary}[section]

\newcommand{\bp}{\begin{Proposition}}	\newcommand{\ep}{\end{Proposition}}
\newcommand{\bt}{\begin{Theorem}}	\newcommand{\et}{\end{Theorem}}
\newcommand{\bl}{\begin{Lemma}}		\newcommand{\el}{\end{Lemma}}
\newcommand{\bc}{\begin{Corrolary}}	\newcommand{\ec}{\end{Corrolary}}

\newcommand{\bra}{\langle}
\newcommand{\ket}{\rangle}
\def\arXiv#1{\href{http://arxiv.org/abs/#1}{arXiv:#1}}
\def\arXiv#1#2{\href{http://arxiv.org/abs/#1}{arXiv:#1}}

\def\doi#1#2{\href{http://doi.org/#1}{#2}}

\begin{document}

\title{Topological modes in relativistic hydrodynamics}

\author{Yan Liu$^{a,b}$}\email{yanliu@buaa.edu.cn}
\author{Ya-Wen Sun$^{c,d}$}\email{yawen.sun@ucas.ac.cn}

\affiliation{\vspace{.2cm}${}^a$Center for Gravitational Physics, Department of Space Science, Beihang University, Beijing 100191, China\\
${}^b$Key Laboratory of Space Environment Monitoring and Information Processing, 
 Ministry of Industry and Information Technology, Beijing 100191, China\\
${}^c$School of physics $\&$ CAS Center for Excellence in Topological Quantum Computation, University of Chinese Academy of Sciences, Beijing 100049, China\\
 ${}^d$Kavli Institute for Theoretical Sciences,  University of Chinese Academy of Sciences, Beijing 100049, China}

\begin{abstract}
We show that gapless modes in relativistic hydrodynamics could become topologically nontrivial by weakly breaking the conservation of energy momentum tensor in a specific way. This system has topological semimetal-like crossing nodes in the spectrum of hydrodynamic modes that require the protection of a special combination of translational and boost symmetries in two spatial directions. 
We confirm the nontrivial topology from the existence of an undetermined Berry phase. These energy momentum non-conservation terms could naturally be produced by an external gravitational field that comes from a reference frame change from the original inertial frame, i.e. by fictitious forces in a non-inertial reference frame. This non-inertial frame is the rest frame of an accelerating observer moving along a trajectory of a helix. This suggests that topologically trivial modes could become nontrivial by being observed 
in a special non-inertial reference frame, and this fact could be verified in laboratories, in principle. Finally,  we propose a holographic realization of this system.

\end{abstract}%
\maketitle

\section{ Introduction} 

Hydrodynamics is the universal low energy theory for systems close to local thermal  equilibrium at a long distance and time. It could describe a variety of physical systems ranging from matter at large scales in the Universe, the quark-gluon plasma \cite{CasalderreySolana:2011us}, to Weyl semimetals \cite{Landsteiner:2014vua,Lucas:2016omy} and graphenes \cite{Lucas:2017idv} in the laboratory. At small momentum and frequency, perturbations of a hydrodynamic system close to the equilibrium would produce propagating as well as diffusive modes \cite{Kovtun:2012rj}. These modes are gapless,  whose poles are at $\omega={\bf k}=0$, which reflects the fact that energy momentum is conserved. 

During the last decade, topologically nontrivial quantum states have been discovered in condensed matter physics \cite{vishwanath,Witten:2015aoa}. Later it has been found that many classical systems have nontrivial topological states too, including topological optical/sound systems (see e.g. \cite{topphoton, topphoton2, natphy} and references therein), which have also been observed experimentally. 

It raises the question if the gapless modes in relativistic hydrodynamics could also become topologically nontrivial under certain conditions. In this paper, we start from the relativistic hydrodynamics and show that after weakly breaking conservation of energy momentum, hydrodynamic modes could become topological semimetal-like nontrivial states that require  the protection of a special spacetime symmetry, and interestingly, these non-conservation terms for the energy momentum tensor could come from a non-inertial reference frame of an accelerating observer moving along a helix.
\section{Effective Hamiltonian and spectrum in relativistic hydrodynamics}
 We focus on the simplest hydrodynamic systems with no internal charges whose only conserved quantity is the energy momentum tensor that satisfies
$
\partial_\mu T^{\mu\nu}=0\,.
$
Up to the first order in derivative, the constitutive equation for the energy momentum tensor in the Landau frame is
\bea
T^{\mu\nu}&=\epsilon u^\mu u^\nu+P \Delta^{\mu\nu}-\eta \Delta^{\mu\alpha}\Delta^{\nu\beta}\big(\partial_\alpha u_\beta +\partial_\beta u_\alpha 
\nn\\
&
-\frac{2}{3}\eta_{\alpha\beta}\partial_\sigma u^\sigma\big)-\zeta \Delta^{\mu\nu}\partial_\alpha u^\alpha+\mathcal{O}(\partial^2)\,,\nn 
\eea  
where $\Delta_{\mu\nu}=\eta_{\mu\nu}+u_\mu u_\nu$, $\epsilon$, $P$ are the energy densities and pressure and $\eta$, $\zeta$ are the shear and bulk viscosities.

With small perturbations away from the equilibrium, the system would respond to the perturbations and develop hydrodynamic modes.  
There are four eigenmodes of the system. Two of them are the sound modes propagating in the direction of ${\bf k}=(k_x,k_y,k_z)$ with the dispersion relation $\omega=\pm v_s k-i\Gamma_s k^2$
, where $ v_s=\sqrt{\frac{\partial P}{\partial \epsilon}}$ and $\Gamma_s=(\frac{4}{3}\eta+\zeta)/(\epsilon+P)$. 
The other two are transverse modes with $\omega=-i\frac{\eta}{\epsilon+P} k^2$. 
To the first order in $k$, dissipative terms disappear and the spectrums of the four modes are real, which cross each other at $\omega={\bf k}=0$.
This spectrum looks similar to the spectrum of Dirac semimetals, except that we have two extra flat bands here. 

To change the spectrum to a topological semimetal-like one, we need to add non-conservation terms of $T^{\mu\nu}$ into the conservation equations.
As a first step, we develop the notion of an effective Hamiltonian in hydrodynamics. Substituting the constitutive equations for the perturbations $\delta T^{\mu\nu}$ into 
$\partial_\mu \delta T^{\mu\nu}=0\,$, 
we could rewrite the equations into the form
\be
i\partial_t\Psi=H\Psi
\ee
where we have defined
\be\label{effH}
\Psi=\begin{pmatrix} 
\delta \epsilon \\
\delta \pi^x \\
\delta \pi^y  \\
\delta \pi^z
\end{pmatrix}\,,~~~~~~
H=\begin{pmatrix} 
0 & ~~k_x & ~~k_y & ~~ k_z \\
k_x v_{s}^2 & ~~0 & ~~ 0 & ~~ 0 \\
k_yv_{s}^2 & ~~ 0 & ~~ 0 & ~~0 \\
k_z v_{s}^2 & ~~0 & ~~ 0 & ~~ 0 
\end{pmatrix}\,
\ee at leading order in $k$, {\em i.e.} omitting dissipative terms at $\mathcal{O}(k^2)$.

In this way, in analogy to the electronic systems \cite{shenbook} we have defined an effective Hamiltonian matrix $H$, whose eigenvalues give the 
spectrum of hydrodynamic modes \cite{footnote1}. 
The four eigenvalues of the matrix Hamiltonian above give the sound modes $\omega=\pm v_s \sqrt{k_x^2+k_y^2+k_z^2}$ and double copies of transverse modes $\omega=0$. The form (\ref{effH}) is the ``free" Hamiltonian matrix for a conserved energy momentum tensor.

\section{Topologically nontrivial modes} 
To deform the spectrum of the hydrodynamic modes, we introduce non-conservation terms for the energy momentum tensor and make sure that the non-conservation terms are small enough to stay within the hydrodynamic limit. The non-conservation of energy and momentum could come from a certain external system, which couples to the hydrodynamic system under study. At this stage we assume that the constitutive equations for $T^{\mu\nu}$ do not get modified, and later, we will take the modifications into account and show that the spectrum does not change up to a rescaling of parameters.

We take a 4D hydrodynamic system and introduce non-conservation terms for $T^{\mu\nu}$ as follows
\bea\label{eq:4dhydro}
\begin{split}
\partial_\mu \delta T^{\mu t}&=m \delta T^{tx}\,,~~
\partial_\mu \delta T^{\mu x}=-m v_s^2 \delta T^{tt}
\,,\\
\partial_\mu \delta T^{\mu y}&=b v_s \delta T^{tz}\,,~~
\partial_\mu \delta T^{\mu z} =-b v_s \delta T^{ty}\,,
\end{split}
\eea
where $m$ terms gap the spectrum, while $b$ terms change the momentum position of the crossing nodes in the spectrum and we assume $\mathcal{O}(k^2) < \mathcal{O}(m,b)\lesssim \mathcal{O}(k)$. 

Physically, (\ref{eq:4dhydro}) states that energy (momentum in the $x$ direction) is not conserved,  whose non-conservation is proportional to the momentum in the $x$ direction (energy). Later, we will show that these seemingly {\it ad hoc} non-conservation terms naturally arise from the observation of an accelerating observer moving in a helix, which could be tested in experiments in principle. 

After substituting the fluctuations of constitutive equation into (\ref{eq:4dhydro}) we obtain  
\be
\begin{split}
i\partial_t\Psi=H\Psi
\end{split}
\ee
where $\Psi=\big(\delta \epsilon, \delta \pi^x, \delta \pi^y, \delta \pi^z\big)^T$ and 
\vspace{-0.2cm}
\be\label{h4d}
H=\begin{pmatrix} 
0 & ~~k_x+im & ~~k_y & ~~ k_z \\
(k_x-im)v_{s}^2 & ~~0 & ~~ 0 & ~~ 0 \\
k_yv_{s}^2 & ~~ 0 & ~~ 0 & ~~ib v_s \\
k_z v_{s}^2 & ~~0 & ~~ -ib v_s & ~~ 0 
\end{pmatrix}\,.
\ee
$H$ is similar to a Hermitian matrix as could be seen by redefining $\delta \epsilon \to \frac{1}{v_s}\delta \epsilon$.
Thus, this effective $H$ has real eigenvalues
and the factor $v_s$ could be ignored which could be  taken back by an inverse transformation when necessary. 

The spectrum of the hydrodynamic modes for (\ref{h4d}), 
\begin{widetext}
\begin{align} 
\label{spectrum1}
\omega=\pm \frac{1}{\sqrt{2}}\sqrt{b^2 + k^2  + m^2\pm\sqrt{(k_x^2 + m^2 - b^2)^2+(k_y^2 + k_z^2)^2 + 2 (k_y^2 + k_z^2) (k_x^2 + m^2 + b^2)}}\,,\nn
\end{align} 
\end{widetext}
where $k=\sqrt{k_x^2+k_y^2 + k_z^2}$.

\begin{figure}[h!]
  \centering
  \includegraphics[width=0.240\textwidth]{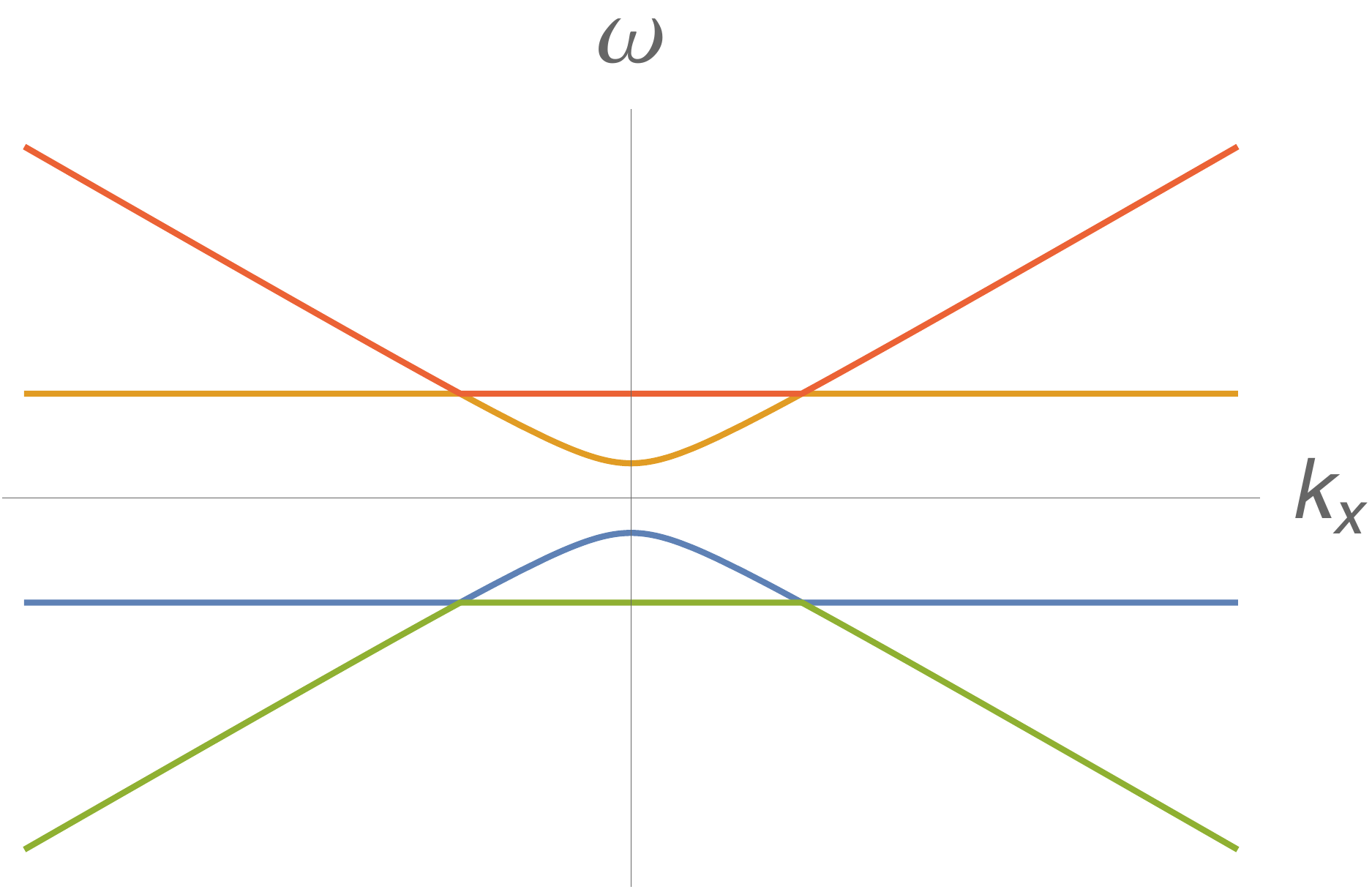}~~
    \includegraphics[width=0.240\textwidth]{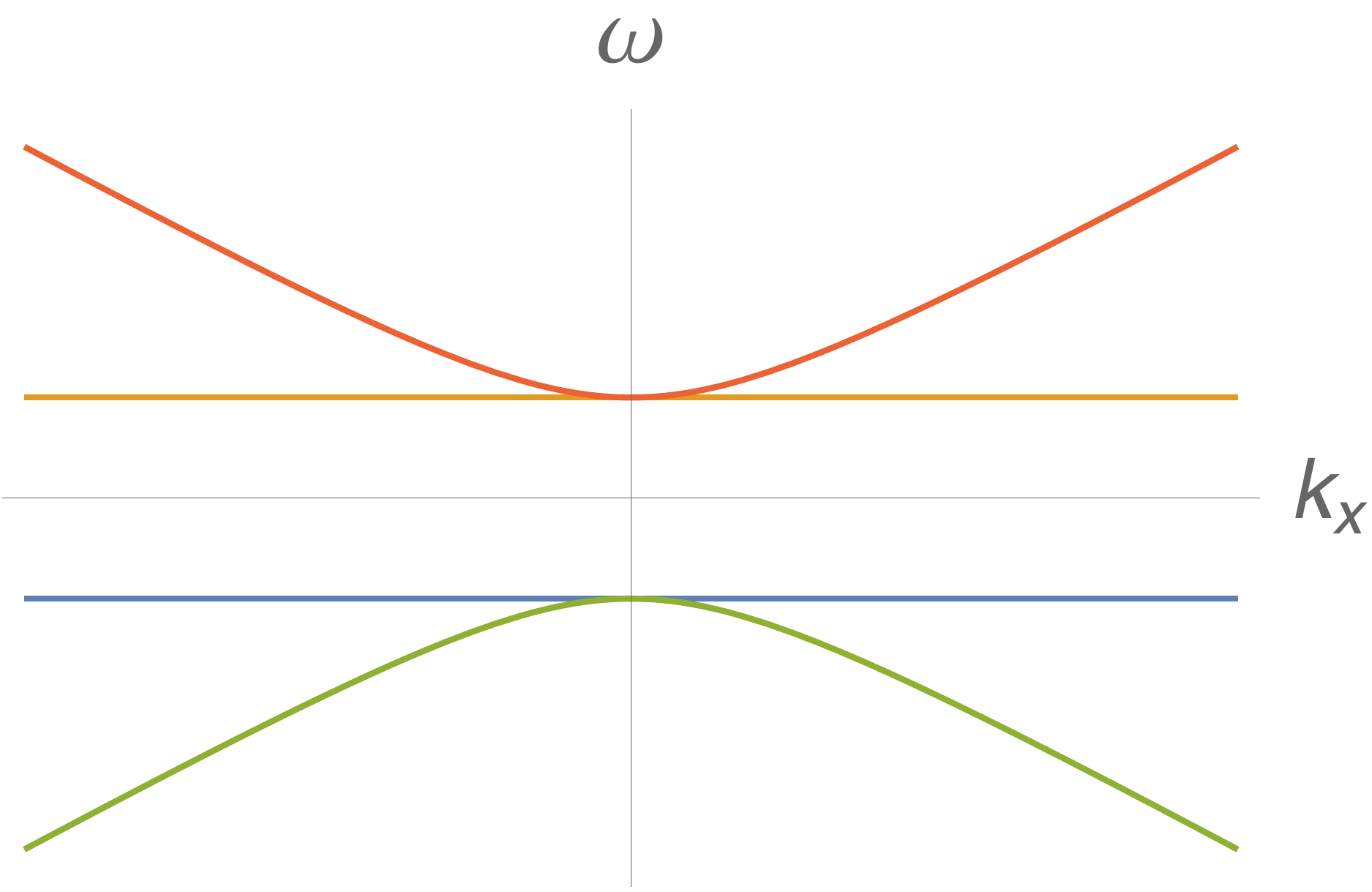}\\~~
      \includegraphics[width=0.240\textwidth]{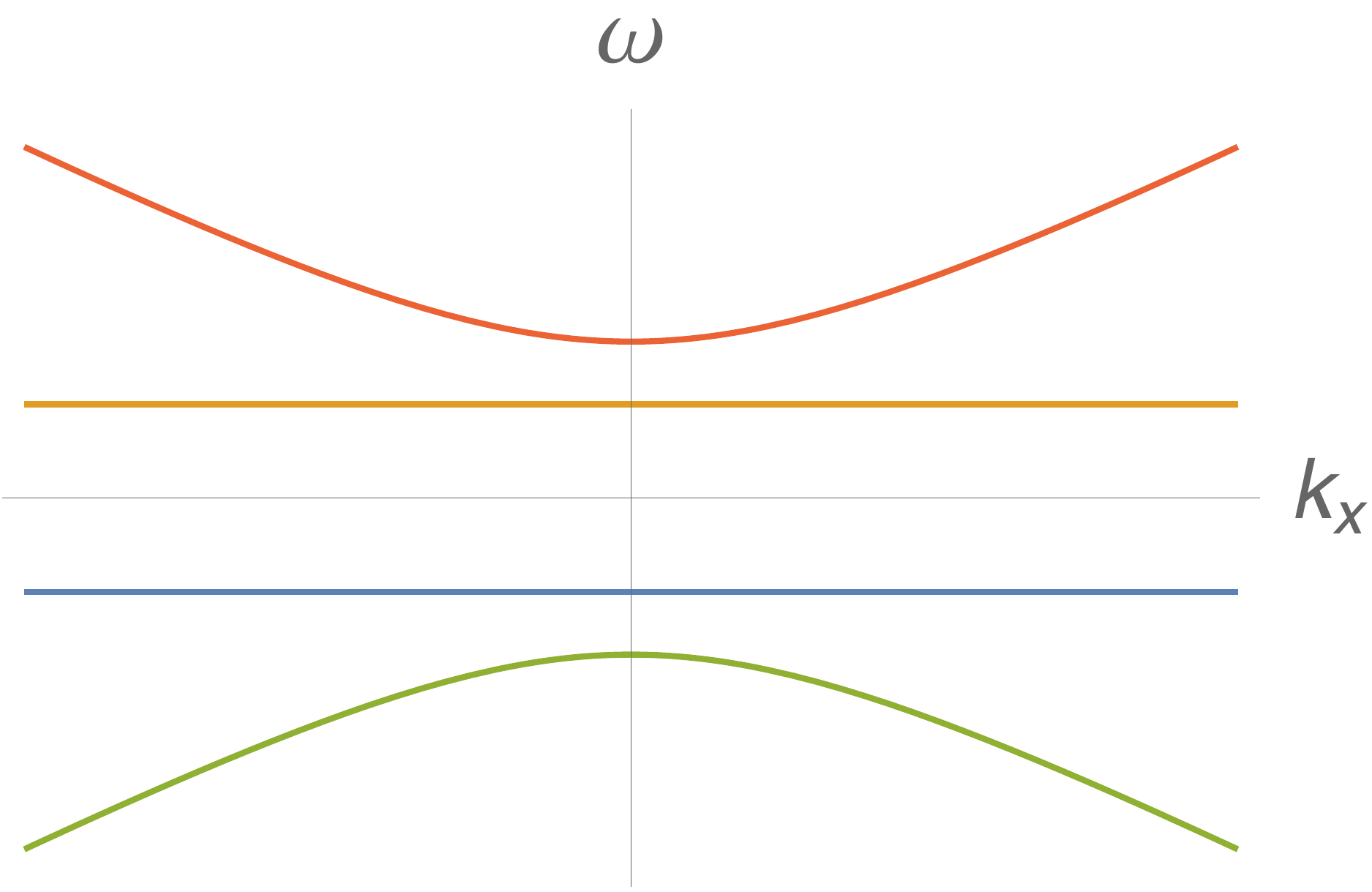}~~
        \includegraphics[width=0.240\textwidth]{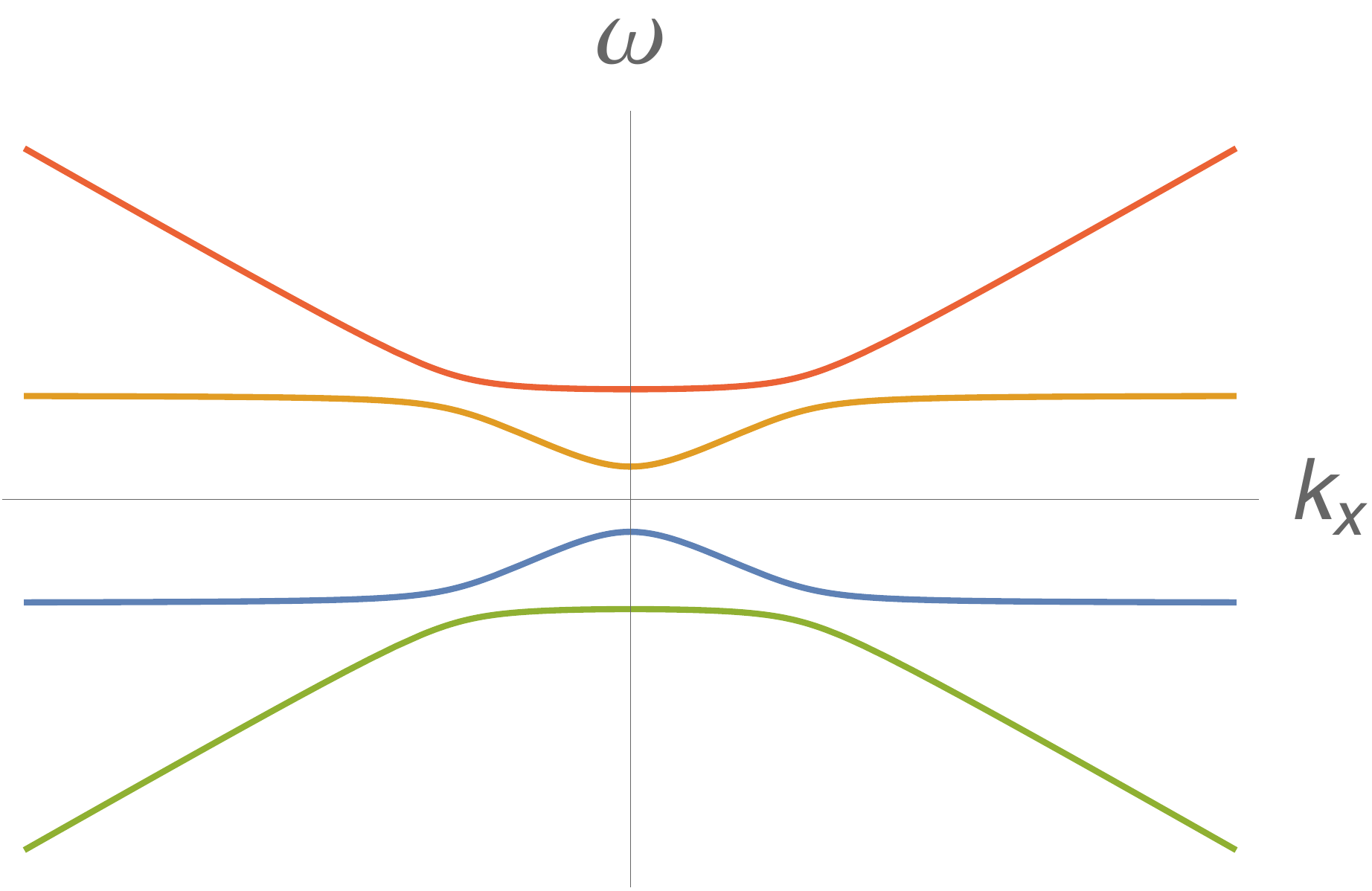}
  \caption{\small The spectrum of the modified hydrodynamics with dynamical equation (\ref{eq:4dhydro}). From left up to right down: the first three plots are for $m<b$, $m=b$, and $m>b$,  respectively, and $k_y=k_z=0$. The fourth plot is for $k_y> 0$, $k_z=0$, and  $m<b$. The distance between two flat bands is determined by the value of $b$ while the gap between two curved bands is determined by the parameter $m$.}
  \label{fig:4D1}
\end{figure}

Figure \ref{fig:4D1} shows this spectrum as a function of $k_x$ for $k_y=k_z=0$ in three different situations: $m<b$, $m=b$ and $m>b$ as well as for $k_y> 0, k_z=0$ at  $m<b$. The effect of $m$ terms is to gap the two sound modes. The effect of $b$ terms is to lift and lower the two transverse flat bands to symmetric positions of opposite sides of the $k$ axis. In this way, the modes have band crossings at nonzero values of $k$ for $m<b$.

From Fig. \ref{fig:4D1}, we could see that for $m<b$,  there are four band crossing nodes at $k_y=k_z=0$ while $k_x \neq 0$, and for these nodes, $\omega\neq 0$. 
These four nodes are still points in the expanded space of $\omega$, $k_x$, $k_y$, and $k_z$ as can be seen from the fourth plot in Fig. \ref{fig:4D1}.  
For $m=b>0$, the system becomes critical with two  nodes, and for $m>b$, the system becomes gapped again. This behavior is qualitatively similar to the topological phase transition of a topological semimetal \cite{Landsteiner:2015pdh}.  

The $m$ terms in (\ref{h4d}) do not gap the four band crossing nodes in the $m<b$ case; however, if we have  extra $m$ terms in the $y$ or $z$ directions, the gaps will open no matter how small the $y$ or $z$ mass parameters are. The spectrum in this case looks the same as the bottom right one in Fig. \ref{fig:4D1}. In this situation, there are no crossing nodes anymore. 
This means that the nodes should be topologically nontrivial under the protection of symmetries that forbid the $m$ terms in the $y$ and $z$ directions. We will see later that the symmetry needed here is a special combination of translational and boost symmetry in $y$ and $z$ directions. In this sense, the system (\ref{eq:4dhydro}) experiences a symmetry protected topological phase transition that happens at the critical point $m=b$. 

Note that for the hydrodynamical modes, $\omega(k=0)$ is not zero anymore due to the non-conservation of energy, i.e. energy is constantly pumped into or out of the system.  
These crossing nodes at $m<b$ are dissipative when order $k^2$ terms are taken into account. This is different from the $\omega=k=0$ nodes which are real poles in hydrodynamics with unbroken translational symmetries.

The new $m$ and $b$ terms above are not dissipative so they only change the shape of the spectrum while do not introduce any imaginary parts in the dispersion relation. In contrast, momentum dissipation terms in, e.g., \cite{Hartnoll:2007ih, Hartnoll:2016apf, Grozdanov:2018fic} are dissipative terms.
%
\section{Origin for non-conservation terms of $T^{\mu\nu}$ }
 The simplest way to have the non-conservation terms of $T^{\mu\nu}$ in (\ref{eq:4dhydro}) is to introduce an external rank two symmetric tensor field. A natural possibility is for this external field to be a gravitational field $h_{\mu\nu}$ \cite{footnote-fmunu}. Then the whole metric field is $g_{\mu\nu}=\eta_{\mu\nu}+h_{\mu\nu}$, and the energy momentum tensor is conserved as $\nabla_\mu  T^{\mu\nu}=0$ in the new spacetime so that $\partial_\mu  T^{\mu\nu}=0$ does not hold anymore. Expanding this equation in $h_{\mu\nu}$, we get
\be\label{eq:conhmunu}
\partial_\mu\delta T^{\mu\nu}=-\frac{1}{2}\partial_\alpha h \delta T^{\alpha\nu}-\frac{1}{2}\eta^{\nu\beta}(2\partial_{\mu}h_{\alpha\beta}-\partial_\beta h_{\mu\alpha})\delta T^{\mu\alpha}\,.
\ee Again, we  have assumed that $\mathcal{O}(h_{\mu\nu})\sim \mathcal{O}(k)$ and only kept leading order in $k$ terms \cite{footnote2}. To get the exact $m$ and $b$ terms in the effective Hamiltonian (\ref{h4d}), there are infinite many choices for $h_{\mu\nu}$ and the simplest choice is for $h_{\mu\nu}$ to be \cite{footnote3}
\bea\label{eq:fvalue}
 \begin{split}
&h_{tt}=h_{xx}=m x\,,~~h_{tx}=h_{xt}=\frac{1}{2}m t(v_s^2+1)\,,~~\\
&h_{ty}=h_{yt}=-\frac{1}{2}b v_s z\,,~~h_{tz}=h_{zt}=\frac{1}{2}b v_s y\,.
\end{split}
\eea

This graviton field $h_{\mu\nu}$ (\ref{eq:fvalue}) could come from sources of massive matter, and more interestingly, it could also come from a reference frame transformation from the flat Minkowski metric generated by $\tilde{x}_{\mu}=x_{\mu}+\xi_{\mu}$ with
\cite{footnote4},  
\be\label{eq:diff}
\xi_\mu=\bigg(\frac{mxt}{2}, ~~\frac{mx^2}{4}+\frac{mt^2}{4}v_s^2,~~-\frac{b}{4}v_s z t,~~\frac{b}{4}v_s y t\bigg)\,.\ee  
This is an intriguing result as usually a nontrivial gravitational field could not be transformed to a flat spacetime globally but only locally. It could be checked that this new metric field has all the components of the Riemann tensor vanishing at leading order, thus could be transformed to the flat spacetime. Though equivalent to a flat spacetime, $h_{\mu\nu}$ could still be viewed as a nontrivial gravitational field according to the equivalence principle. This $h_{\mu\nu}$ denotes a non-inertial reference frame. {\it This result suggests that in a specific non-inertial frame, we could observe hydrodynamic modes that are topologically protected even when they are topologically trivial in the original inertial frame. This could be tested in laboratories, in principle.}

Note that with a nonzero $h_{\mu\nu}$, the constitutive equations for $T_{\mu\nu}$ could also be written into a covariant form thus leading to extra terms compared to the original constitutive equations. However, it can be explicitly checked that these extra terms do not change the spectrum (\ref{eq:4dhydro}) at all or do not change the spectrum up to a rescaling of parameters $m$, $b$, and $v_s$ depending on whether the fluid is resting in the original inertial frame or is accelerating together with the accelerating observer; i.e., the new spectrum could be obtained by substituting the rescaling relation above to the spectrum (\ref{eq:4dhydro}), and this does not mean that the real speed of sound changes. More details could be found in the appendix.
\section{The exact non-inertial reference frame} 
We could work out the exact reference frame from the infinitesimal transformation (\ref{eq:diff}). A rest observer in the new frame $\tilde{x}^{\mu}$ has $d \tilde{x}^{i}=0$ for $i=1,2,3$. From this we could obtain $d\tilde{t}=dt$, $dx=-\frac{m v_s^2 t dt}{2}$, $dy=\frac{b v_s z dt}{4}$, and $dz=-\frac{b v_s y dt}{4}$ at order $O(m)$. From the last two equations above, we have $y=R_0 \cos\frac{b v_s}{4} t  $ and $z=-R_0\sin\frac{b v_s}{4} t  $  with appropriate choice for $t=0$ and a constant radius $R_0$. These together confirm that the rest observer of the new non-inertial frame is in fact an accelerating observer in the original inertial frame who has a constant acceleration $a=-\frac{m v_s^2}{2}$ in the $x$ direction and a constant angular velocity $\omega_x=\frac{b v_s}{4} $ in the $y$ and $z$ plane. As $m$ and $b$ are small parameters, the observer is moving in the non-relativistic limit, which is consistent with $d\tilde{t}=dt$. Note that the fluid is still a relativistic one though moving collectively non-relativisitically.

\begin{figure}[h!]
  \centering
 \vspace{-1.6 cm}
  \includegraphics[width=0.440\textwidth]{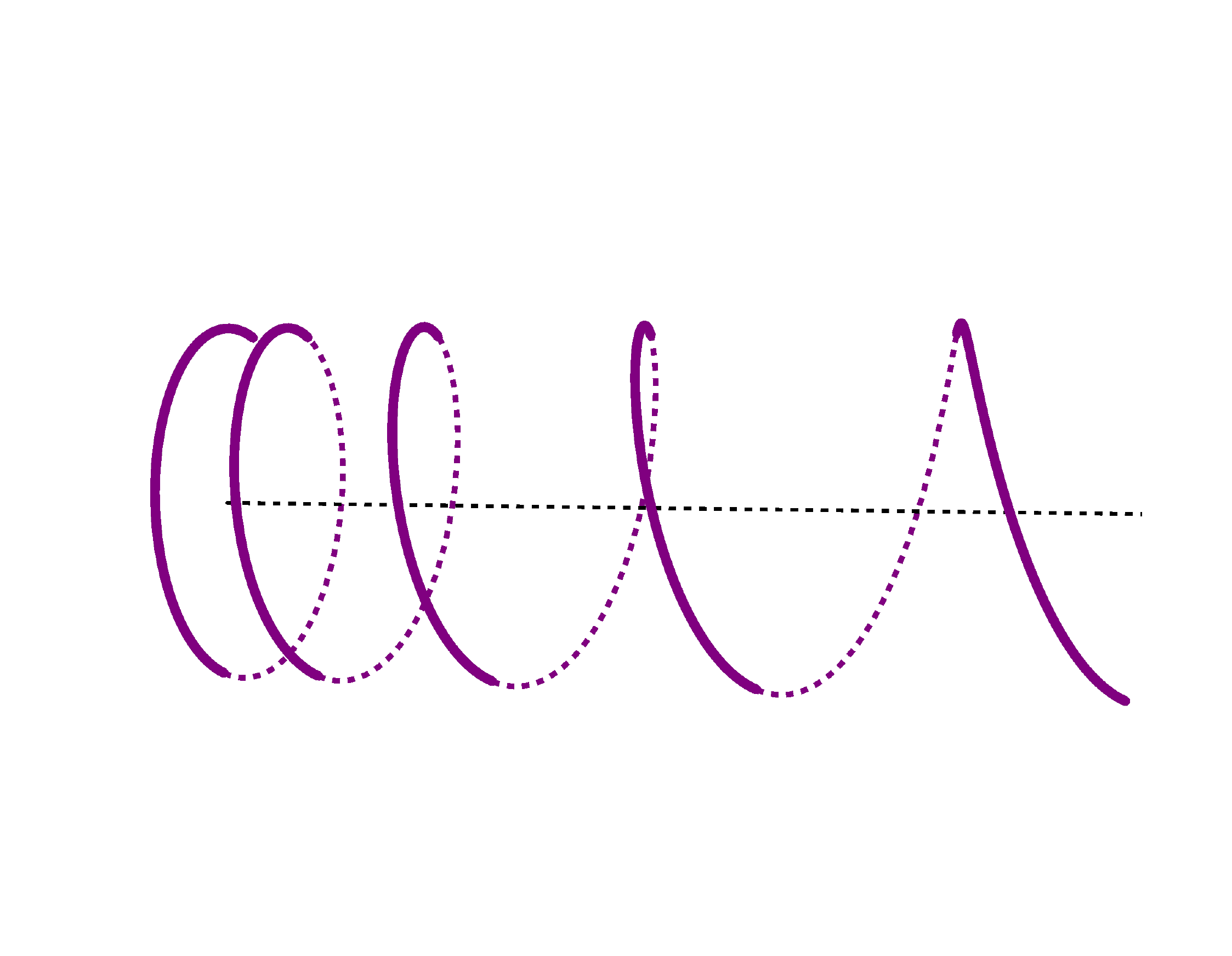}
\vspace{-1.6cm}
  \caption{\small The helix trajectory of the accelerating observer and the fluid in the laboratory frame. The observer/fluid has both a constant acceleration and a constant angular velocity in the $x$ direction. }
  \label{fig:helix}
\end{figure}

Thus, the topological modes are those observed by an accelerating observer moving together with the fluid along a helix in the non-relativistic limit as shown in Fig. \ref{fig:helix}, which makes it, in principle, a realizable setting for experimental tests of this system. 
Physically, the non-conservation terms for $T^{\mu\nu}$ could be thought of as coming from the fictitious force, including the Coriolis force, the  centrifugal force as well as the inertial force associated with the $x$ direction constant acceleration. The non-trivial topological nodes could also be viewed as coming from these fictitious effects. Now we have shown that the seemingly {\it ad hoc}  non-conservation terms of $T^{\mu\nu}$ could in fact be generated from a very natural non-inertial reference frame.

Now we could work out the symmetry of the system (\ref{eq:4dhydro}) as the isometry of the metric $g_{\mu\nu}=\eta_{\mu\nu}+h_{\mu\nu}$, i.e. coordinate transformations that leave $g_{\mu\nu}$  unchanged. The symmetry could be viewed as the Lie transformation of the Poincare symmetry generated by the vector (\ref{eq:diff}). Among the ten generators of this new isometry \cite{Liu:2020abb}, two of them are responsible for forbidding $m$ terms in the $y$ and $z$ directions and protecting the nontrivial topological states, which are generated by $x^{\mu}\to x^{\mu}+\epsilon^{\mu}$, where $\epsilon^{\mu}=a_y \chi_y+a_z \chi_z$ with $ \chi_y=\left(-\frac{b z v_s}{4}, ~0, ~1, ~-\frac{b t v_s}{4}\right)$, $\chi_z=\left(\frac{b  yv_s}{4}, ~0, ~\frac{b yv_s}{4}, ~1\right)$ and $a_y$, $a_z$ two infinitesimal constants. $\chi_y$ (or $\chi_z$) is a special combination of the $y$ (or $z$) direction translational symmetry and the boost symmetry of the $t$-$y$ (or $t$-$z$) direction. Though this symmetry looks complicated, physically it only requires the covariant conservation of momentum in the $y$ and $z$ directions in the non-inertial frame; i.e. there are no extra external forces in the $y$ and $z$ directions.

Finally we mention another possible circumstance to have this nonzero $h_{\mu\nu}$, which could arise in analog gravity systems, i.e. certain materials could give rise to effective hydrodynamic equations as if there exists a nontrivial gravitational field. 
\section{Topological invariant}
For systems protected by a certain symmetry, we could calculate the topological invariant at a high symmetric point in the momentum space, which is $k_y=k_z=0$ in this case. There is a charge conjugation symmetry for the solutions and we could focus on the lower two nodes in the left top plot of Fig.  \ref{fig:4D1}. 
Here, as we are in zero effective residual spatial dimension,  the calculation of the topological invariant is different from the Berry phase or Berry curvature for a nodal line or Weyl semimetals. For the left node at $k_x=k_1$ in the left top plot of Fig. \ref{fig:4D1}, the green solution at the left limit $k_x\to k_{1-}$ and the right limit $k_x\to k_{1+}$ are denoted as $| n_1\ket$ and 
$|n_2\ket$ separately.   
We could define a Berry phase between the two states $e^{-i \alpha}=\frac{\bra n_1|n_2\ket}{|\bra n_1|n_2\ket |}$ to denote the topological invariant here. If the Berry phase is an undetermined one; i.e., $| n_1\ket$ and 
$|n_2\ket$ are orthogonal to each other, 
the system would be topologically nontrivial as the two states cannot be connected without passing through a singularity, which means the lower band and the upper band could not be separated by small perturbations.

In (\ref{eq:4dhydro}), $|n_1\ket=\frac{1}{\sqrt{2}}\big(0,0,-i,1\big)$ and $|n_2\ket=\frac{1}{\sqrt{1+\frac{1}{v_s^2}}}\Big(-\frac{i\sqrt{m^2+k_x^2}}{v_s(m+i k_x)},1,0,0\Big)$. Therefore $\bra n_1|n_2\ket=0$, which means that the Berry phase is undetermined. From the argument above the two bands cannot be separated easily by a gap without going through a topological phase transition. Similar behavior of an undetermined Berry phase has also happened for the holographic nodal line semimetals \cite{Liu:2018djq, Landsteiner:2019kxb}. 

This result confirms that the four nodes in Fig.  \ref{fig:4D1} are topologically nontrivial protected by a special combination of translational and boost symmetry in the $y$ and $z$ directions. At the same time, the Berry phase accumulated through the whole circle around this node would be trivial indicating that it is indeed topologically trivial without the symmetry. 
\section{Transport properties} 
We can follow the calculations in \cite{Kovtun:2012rj, Davison:2014lua} to compute the heat transport for this system to uncover more observational effects.  We obtain
\bea
\kappa_{xx}(\omega, k_x)&=&-\frac{i\omega(\epsilon+P)}{T\left((k_x^2+m^2)v_s^2+i\frac{\eta}{\epsilon+P}\omega k_x^2-\omega^2\right)}\,,\nn\\
\kappa_{yy}(\omega, k_x)&=&\kappa_{zz}(\omega, k_x)=
-\frac{k_x^2\eta+i\omega(\epsilon+P)}{T\left(b^2 v_s^2+(i\omega+\frac{\eta}{\epsilon+P}k_x^2)^2\right)}\,,\nn\\
\kappa_{yz}(\omega, k_x)&=&-\kappa_{zy}(\omega, k_x)=\frac{ (\epsilon+P)b v_s}{T\left(b^2 v_s^2+(i\omega+\frac{\eta}{\epsilon+P}k_x^2)^2\right)}\,.\nn
\eea

With the formulas above, when $m=b=0$, all diagonal components of the dc heat transport diverge. For generic $m$ and $b$, we have vanishing dc heat transport $\kappa_{xx}(0,0), \kappa_{yy}(0,0)$ and 
$\kappa_{zz}(0,0)$ while $\kappa_{yz}(0,0)=-\kappa_{zy}(0,0)=\frac{\epsilon+P}{Tbv_s}.$ These $m$ and $b$ terms eliminate the unphysical divergence of dc heat transports and lead to interesting vanishing dc heat transport behavior. An intuitive and physical reason for this system to be a dc  thermal insulator could be seen from the spectrum, where there is an energy gap at $k=0$ in contrast to being gapless at $k=0$ for standard hydrodynamics, as the consequence of energy non-conservation. This is similar to the mechanism of finite dc conductivity obtained for cases with momentum dissipation. 
%
\section{$\mathcal{O}(k^2)$ effects}
$\mathcal{O}(k^2)$ terms lead to dissipative effects and give rise to imaginary parts of frequency in the spectrum. 
The $\mathcal{O}(k^2)$ terms 
make the effective Hamiltonian matrix  non-Hermitian. Here, we still keep terms at $m\sim b$ order while not $m^2\sim b^2$ order assuming that $m\sim k^2$.

From the eigenvalues of Hamiltonian with $\mathcal{O}(k^2)$ effects included, we find that the real part has not  changed while imaginary parts appear. At the four nodes, the imaginary parts are not zero indicating that the four nodes are dissipative in comparison to nondissipative nodes at $\omega=0$ in the usual hydrodynamics. 
The imaginary part for each of the band has a jump at the crossing nodes at $k_y=0$ in the $k_x$ axis, i.e. the imaginary parts of the same band are different at the left and right limits of the singular node. This behavior is similar to the behavior of the eigenstates when calculating the Berry phase and thus 
 provides another piece of evidence of the existence of a symmetry protected topological singular node.
\section{Ward identities and holographic realization} 
The physics of hydrodynamics has been studied extensively in holography for strongly coupled systems  \cite{Kovtun:2004de, Son:2007vk,Rangamani:2009xk}. We aim to construct a holographic system possessing  the same non-conservation equation of (\ref{eq:4dhydro}), thus providing an example of this system in the strongly coupled limit. 

 Holographically we could also perform a coordinate transformation to get a non-inertial frame version of AdS/CFT correspondence which has the metric $g^{\mu\nu}=\eta_{\mu\nu}+h_{\mu\nu}$ at the boundary. This system should have the same spectrum of the hydrodynamic modes. As a first step for a confirmation, we need to show that our holographic system indeed has the non-conservation of (\ref{eq:4dhydro}). For this purpose, we will first obtain the Ward identities for $T^{\mu\nu}$ in the non-conserved hydrodynamic system from (\ref{eq:4dhydro}) and match these identities to those  in the holographic non-inertial frame system. %

In the case that these non-conservation terms come from a gravitational field, we could start from the covariant conservation equation $\nabla_{\mu}T^{\mu\nu}=0$ and differentiate it with respect to $ g_{\lambda\rho}$ to obtain the Ward identities in the  momentum space of the boundary system. 
To the first order in $h_{\mu\nu}$, the Ward identities are
\bea\label{eq:ward}
&k_\mu G^{\mu\nu,\lambda\rho}(k)+i\Big[\Gamma^{(1)\mu}_{~~~~\mu\alpha}G^{\alpha\nu,\lambda\rho}(k)+\Gamma^{(1)\nu}_{~~~~\mu\alpha}G^{\mu\alpha,\lambda\rho}(k)\Big]\nn\\
&~~~+\text{contact terms}=0\,,
\eea where the explicit form of the contact terms is omitted. With nonzero $h_{\mu\nu}$ several components of $\Gamma^{(1)\nu}_{~~~~\mu\alpha}$ would be nonzero and contribute extra terms to the Ward identities of $k_\mu G^{\mu\nu,\lambda\rho}(k)+\text{contact terms}=0$ in hydrodynamics systems with conserved $T^{\mu\nu}$ \cite{Policastro:2002tn}.

 The Ward identities (\ref{eq:ward}) could be reproduced from the holographic non-inertial frame system where we start from the usual AdS Schwartzchild black hole and perform coordinate transformations so that the boundary metric becomes $g_{\mu\nu}=\eta_{\mu\nu}+h_{\mu\nu}$. 
  
We have checked that the holographic Ward identities match exactly to the hydrodynamic Ward identities (\ref{eq:ward}). The details will be presented in \cite{Liu:2020abb}, and the hydrodynamic modes and Green functions will be systematically studied in future work. 
\section{Outlook} 
One important application of the observation of topologically nontrivial hydrodynamics modes at finite frequency would be to enlarge the amplitudes at the crossing frequencies and momenta due to the doubling of modes and this enlargement could be stable from perturbations under certain conditions.

The fact that an accelerating observer moving along a helix would see topological hydrodynamic modes of a hydrodynamic system moving in the same helix could,  in principle, be checked in experiments, including probing the spectrum of the modes and measuring the featured transport properties of the system.

Finally, and most importantly, it is possible that systems other than hydrodynamic systems, e.g.,  electronic/photonic systems, would also become topologically nontrivial being observed in a certain non-inertial frame \cite{footnote-ele}, which would provide another way to obtain topologically nontrivial materials by mechanically accelerating the detector in a laboratory. This brings up a new interesting effect for accelerating observers in addition to the well-known Unruh effect \cite{Unruh:1976db}.

\subsection*{Acknowledgments}
We would like to thank Matteo Baggioli, Rong-Gen Cai, Wen-Bin Pan, Koenraad Schalm, Qing Zhang, anonymous referees and especially Karl Landsteiner for useful discussions. This work is supported by the National Key R\&D Program of China (Grant No. 2018FYA0305800). The work of Y.L. was also supported by the National Natural Science Foundation of China Grant No.11875083. The work of Y.W.S. has also been partly supported by starting grants from University of Chinese Academy of Sciences and Chinese Academy of Sciences, and by the Key Research Program of 
Chinese Academy of Sciences (Grant No. XDPB08-1), 
the Strategic Priority Research Program of Chinese Academy of Sciences, 
Grant No. XDB28000000.



\onecolumngrid

\begin{center}
\textbf{\large }
\end{center}
\pagebreak
\widetext
\begin{center}
\textbf{\large Appendix: Topological modes in relativistic hydrodynamics}
\end{center}
\begingroup
\hypersetup{linkcolor=black}
\endgroup

\setcounter{equation}{0}
\setcounter{figure}{0}
\setcounter{table}{0}
\setcounter{page}{1}
\setcounter{section}{0}
\makeatletter
\renewcommand{\theequation}{S\arabic{equation}}
\renewcommand{\thefigure}{S\arabic{figure}}
\renewcommand{\bibnumfmt}[1]{[S#1]}
\renewcommand{\citenumfont}[1]{S#1}

\section{Another possible origin for the rank two symmetric external tensor field}

In this appendix we show the possibility for the rank two symmetric external tensor field $f_{\mu\nu}$ to be  an external symmetric tensor matter field that couples to the energy momentum tensor of the system and contributes an effective $f_{\mu\nu}T^{\mu\nu}$ term to the Lagrangian of the system. With this extra term the energy momentum of the system will not be conserved as it can be transferred to the external system. 
 
Omitting terms at order of $\mathcal{O}(k^2)$ or higher, we get the non-conservation equation for 
$T^{\mu\nu}$
 \be\label{fmunu} \partial_{\mu}T^{\mu}_{~\nu}=-\frac{1}{2}T^{\rho\mu}(2 \partial_{\mu} f_{\rho\nu}-\partial_{\nu} f_{\rho\mu})\,. \ee 
As a simple realization, we could switch on the following components of $f_{\mu\nu}$ to get the form of effective Hamiltonian (\ref{h4d})  \bea\label{eq:fvalue}
 \begin{split}
&f_{tt}=f_{xx}=m x\,,~~f_{tx}=f_{xt}=\frac{1}{2}m t(v_s^2+1)\,,~~\\
&f_{ty}=f_{yt}=-\frac{1}{2}b v_s z\,,~~f_{tz}=f_{zt}=\frac{1}{2}b v_s y\,.
\end{split}
\eea
\vspace{0.1cm}\\

\section{Change in the constitutive equations}

In this appendix we show explicitly that after carefully considering the modifications of the constitutive equations to the equations (\ref{eq:4dhydro}) in the main text, the spectrum does not change qualitatively and the discussions in the main text at the leading order in $k$ stay the same. Note that there are two possibilities for the hydrodynamic system: accelerating in the same way as the observer and accelerating in a different way. 

We will first consider the case where the observer is in a non-inertial frame and the fluid is at rest in the same non-inertial frame. Then we also consider a second possibility where the observer is accelerating and the fluid is accelerating at another constant acceleration in the original inertial frame, and one special limit would be that the fluid is at rest in the original inertial frame while the observer is accelerating in the helix. For both cases, the spectrum stays the same up to a rescaling of parameters $m$, $b$ and $v_s$, which does not change the qualitative behavior, though there would be a little difference in the explanation of the topological nature. 

\subsection{ I. Observer and fluid accelerating in the same way}
In this part we show explicitly in the case of observer and fluid accelerating in the same way, how the change in the constitutive equations for $T^{\mu\nu}$ due to the graviton field $h^{\mu\nu}$ will not affect the spectrum that was calculated from the constitutive equations in the flat metric except a rescaling of the parameters $m,~b$ and $v_s$. 

For the background metric $g_{\mu\nu}=\eta_{\mu\nu}+h_{\mu\nu}$, the covariant form of the constitutive equation for the energy momentum tensor is \be T^{\mu\nu} =(\epsilon+P)u^{\mu}u^{\nu}+P g^{\mu\nu}\ee at zeroth order in derivative, where $g_{\mu\nu}u^{\mu}u^{\nu}=-1$ and $u^i=0$ for $i=x,y,z$ because the fluid is at rest in the non-inertial frame, i.e. accelerating in the inertial frame. The corrections of $T^{\mu\nu}$ compared to the energy momentum tensor for the flat metric $\eta^{\mu\nu}$ is at order $O(m)\sim O(k)$ as $h^{\mu\nu}$ is of order $O(m)$.

Now we examine carefully how these extra terms contribute to the (non-)conservation equation (\ref{eq:conhmunu}). In (\ref{eq:conhmunu}), the extra terms on the right hand side compared to the original equation are at order $O(k^2)$ thus have no extra contributions compared to original terms when modifications of the constitutive equations were not considered. On the left hand side, in $\partial_\mu \delta T^{\mu\nu} $ extra terms that have derivatives of perturbations $\delta \epsilon,\delta P$ or $\delta u^\mu$ have order $O(k^2)$ thus do not contribute at leading order. Extra terms that might have contributions to the conservation equation are \be \delta (\epsilon+P)\partial_{\mu}(u^\mu u^\nu)+(\epsilon+P)(\delta u^\mu \partial_{\mu}u^{\nu}+\delta u^\nu \partial_{\mu}u^{\mu})+\delta P \partial_{\mu}g^{\mu\nu}.\ee As the only nonvanishing component for $u^\mu u^\nu$ is the $tt$ component, it can be checked that $\partial_{\mu}(u^\mu u^\nu)=0$ and $\delta u^\nu \partial_{\mu}u^{\mu}=0$ as $h^{tt} $ does not depend on $t$. Thus substituting the exact expression for $h^{\mu\nu}$, the nonzero contribution to the (non-)conservation equation (\ref{eq:conhmunu}) only exists for the $\nu=t$ and $\nu=x$ components, each being proportional to the nonconservation term on the right with a different factor in front: for $\nu=t$ there will be a contribution of an extra $-\frac{m}{2} T^{0x}$ on the right and for $\nu=x$ there is an extra $-\frac{m}{2}(1-v_s^2)v_s^2T^{00}$ on the right. Thus these terms only result in a rescaling in the corresponding parameters of $m$ and $v_s$, i.e. the resulting spectrum could be obtained from rescaled parameters $\bar{m},\bar{v_s}, \bar{b}$ in the long expression for the spectrum below (\ref{h4d}) as 
\be \bar{m}=\frac{m}{2}\,,~~ \bar{v_s}=v_s\sqrt{3-v_s^2}\,,~~ \bar{b}=\frac{b}{\sqrt{3-v_s^2}}\,. \ee 
As $v_s$ cannot be larger than $1$ ($v_s=1/\sqrt{3}$ for a conformal field theory and this value is expected to be the upper bound for the speed of sound), all the rescaled parameters above are nonzero thus the rescaling does not change the qualitative behavior of the spectrum. Besides this, there is another similar case with a different set of $h_{\mu\nu}$ which we will discuss in a later work. 

Note that though the fluid is collectively accelerating under an external force $\nabla_{\mu}\delta T^{\mu\nu}=0$ would still hold for perturbations of the fluid slightly away from equilibrium. One could write out the explicit conservation equation of energy momentum of the accelerating fluid with external forces (that accelerate the fluid as seen from the inertial frame) as well as gravitational force on the right side. Then we perturb this system from the equilibrium state and external forces would not contribute to the equations for the perturbations as they stay the same as long as the total mass of the system does not change. Physically this is because the fluid elements only experience internal forces as well as gravitational forces while not external forces except at the boundary of the fluid thus $\nabla_{\mu}\delta T^{\mu\nu}=0$ still holds for perturbations.

\subsection{Observer accelerating and fluid rest in the inertial frame}
Note that from the discussions on the origin of non-conservation terms of $T^{\mu\nu}$ from reference frame changes, we can see that the observer is in a non-inertial frame $g_{\mu\nu}$, and the fluid could stay at rest in the inertial frame. In this case, the hydrodynamic system has the four velocity $u^{\mu}=(-1,~0 , ~0,~0)$. By the reference frame transformation (\ref{eq:diff}) we could obtain the four velocity in the non-inertial reference frame  $u^\mu=(-1-\frac{mx}{2},\frac{m t}{2}v_s^2,-\frac{bv_sz}{4},\frac{bv_s y}{4})$ up to the first order in $m,~b$. In the accelerating frame, $\eta^{\mu\nu}$ in $T^{\mu\nu}$ of the fluid should also be replaced by $g^{\mu\nu}$, which is
\be\label{eq:metric}
g_{\mu\nu}=\begin{pmatrix} 
-1+mx & ~~\frac{mt}{2}(1+v_s^2) & ~~-\frac{1}{2}bv_s z& ~~ \frac{1}{2}b v_s y \\
\frac{mt}{2}(1+v_s^2) & ~~1+m x & ~~ 0 & ~~ 0 \\
-\frac{1}{2}bv_s z & ~~ 0 & ~~ 1 & ~~0 \\
\frac{1}{2}b v_s y & ~~0 & ~~ 0 & ~~ 1 
\end{pmatrix}\,.
\ee

The perturbations $\delta T^{\mu\nu}$ at the leading order are 
\be\label{eq:emfluc}
\delta T^{\mu\nu}= (1+v_s^2)\delta\epsilon u^\mu u^\nu+(\epsilon+P)(\delta u^\mu u^\nu+u^\mu \delta u^\nu)+v_s^2\delta\epsilon g^{\mu\nu}
\ee where $v_s=\sqrt{\frac{\partial P}{\partial \epsilon}}$ and $u^{\mu}$ in this formula is the four velocity of the fluid in the non-inertial frame. 

Note that the four velocities of the fluid is normalized as $g_{\mu\nu}u^\mu u^\nu=-1$, thus one of the fluctuations of these four velocities is not independent. We take the independent fluctuations of the fluid as $\Psi=(\delta\epsilon, \delta\pi^x, \delta\pi^y, \delta\pi^z)^T$ with $\delta\pi^i=-(\epsilon+P)\delta u^i$. Note that here $\delta\pi^x$ is not $\delta T^{tx}$, and the spectrum of $H$ will stay the same if we use instead variables $(\delta\epsilon, \delta u_x, \delta u_y, \delta u_z)^T$ or any other set of combined variables.  Plugging (\ref{eq:emfluc}) into the dynamical equation (\ref{eq:conhmunu}) in the main text, we obtain equation 
$i\partial_t\Psi=H\Psi$ with 
\be
H=\begin{pmatrix} 
0 & ~~k_x-{im} & ~~k_y & ~~ k_z \\
(k_x+ im )v_{s}^2 & ~~0 & ~~ 0 & ~~ 0 \\
k_yv_{s}^2 & ~~ 0 & ~~ 0 & ~~-ib v_s \\
k_z v_{s}^2 & ~~0 & ~~ ib v_s & ~~ 0 
\end{pmatrix}\,
\ee
at leading order in $k$. Thus we get exactly the same spectrum as in (\ref{h4d}). Note that in the calculation, we have assumed that $\delta T^{\mu\nu}(\omega,k)$ could also be expanded as $\omega$ and $k$, and one has to be careful about the Fourier transformation process where certain partial derivatives in $\omega$ and $k$ have to be taken into account.


\begin{thebibliography}{99}
\bibitem{footnote1}
Note that this effective Hamiltonian is different from the hydrodynamic Hamiltonian, e.g. in section 2.4 of \cite{Kovtun:2012rj} in the sense that this effective Hamiltonian matrix $\Psi^T (-k) H \Psi(k)$ gives the ``kinetic" part of the Hamiltonian while the latter only counts in the potential part associated with the source terms.
\bibitem{footnote-fmunu}
Another possibility for $f_{\mu\nu}$ could be found in the appendix.
\bibitem{footnote2}
We assume that we are working in a large but finite volume of spacetime and $m$, $b$ are so small that $m x, b y, b z, mt\ll 1$ in the finite volume. We ignore the boundary effects as the volume is large.
\bibitem{footnote3}
Another general set of choices is to choose $h_{\mu\nu}$ to be 
$h_{tt}=h_{xx}=mx\,,~ h_{tx}=cmt\,,~ h_{ty}=-b v_s z/2\,, h_{tz}=b v_s y/2\,,~ h_{xy}=-my\,,~ h_{xz}=emz\,,$
where the two constants $c$ and $e$ should satisfy $1-2c+(1-2e)v_s^2=-2 v_s^2$.
\bibitem{footnote4}
The set of graviton fields in \cite{footnote3} could also be obtained from a similar coordinate transformation from the flat spacetime.
\bibitem{footnote-ele}
For fermionic systems, we have already had preliminary results showing that a non-inertial frame could also produce topologically nontrivial states from trivial states in the small momentum region.
\end{thebibliography}
\end{document}